\documentclass[final,5p,times,twocolumn]{elsarticle}

\usepackage{epsfig}

\usepackage{amssymb}

\journal{Computer Physics Communications}

\begin{document}

\begin{frontmatter}



\title{An {\it ab} initio study of the magnetic and electronic properties of Fe, Co, and Ni nanowires on Cu(001) surface}


\author[a]{J. C. Tung} \author[a,b]{and G. Y. Guo}
\address[a]{Department of Physics and Center for Theoretical Sciences, National Taiwan University, Taipei 106, Taiwan}
\address[b]{Graduate Institute of Applied Physics, National Chengchi University, Taipei 116, Taiwan}

\begin{abstract}
Magnetism at the nanoscale has been a very active research
area in the past decades, because of its novel fundamental physics and exciting potential applications. We have recently
performed an {\it ab intio} study of the structural, electronic and magnetic properties of all 3$d$ transition metal (TM)
freestanding atomic chains and found that Fe and Ni nanowires have a giant magnetic anisotropy energy (MAE), indicating that
these nanowires would have applications in high density magnetic data storages. In this paper, we perform density functional
calculations for the Fe, Co and Ni linear atomic chains on Cu(001) surface within the generalized gradient approximation, in
order to investigate how the substrates would affect the magnetic properties of the nanowires. We find
that Fe, Co and Ni linear chains on Cu(001) surface still have a stable
or metastable ferromagnetic state. When spin-orbit coupling (SOC) is included, the spin magnetic moments remain almost
unchanged, due to the weakness of SOC in 3$d$ TM chains, whilst significant orbital magnetic moments appear and also are
direction-dependent. Finally, we find that the MAE for Fe, and Co remains large, i.e., being
not much affected by the presence of Cu substrate.
\end{abstract}


\end{frontmatter}


\section{Introduction}
In recent years, nanostructured magnetic materials have received much attention, due to their interesting
physical properties and potential applications. For example, finite free-standing gold atomic chains were first reported in
1998\cite{Ohnishi,Yanson}, and their structural properties, such as the actual
length of the chain have been the focus of intensive experiments and theoretical studies since then.
However, these free-standing atomic chains are unstable and thus can only exist at low temperatures and on
a suitably chosen substrate. Physically stable magnetic nanowires deposited on metallic substrates are
one of the most important nanostructures 
and a variety of techniques
have been used to prepare and study them. For example, Gambardella {\it et al}
\cite{Gambardella1,Gambardella2} succeeded in preparing a high density of parallel atomic chains along steps
by growing Co on a high-purity Pt (997) vicinal surface in a narrow temperature range of 10$\sim$20 K. The magnetism
of the Co wires was also investigated by the x-ray
magnetic circular dichroism.\cite{Gambardella2} Experimentally, copper and tungsten are excellent substrates for growth of Fe thin
films\cite{Hauschild, Tian} because of the small lattice constant mismatch between Fe and Cu (3.61\AA) as well as
W (3.61\AA). Thus, in this paper, we perform first principles calculations to study the electronic and magnetic
properties of the Fe, Co and Ni nanowires. The spin-orbit coupling (SOC) is included in this study to determine the magnetic anisotropy energy (MAE).

\section{Theory and Computational Method}
\begin{figure}
\includegraphics[width=5.5cm]{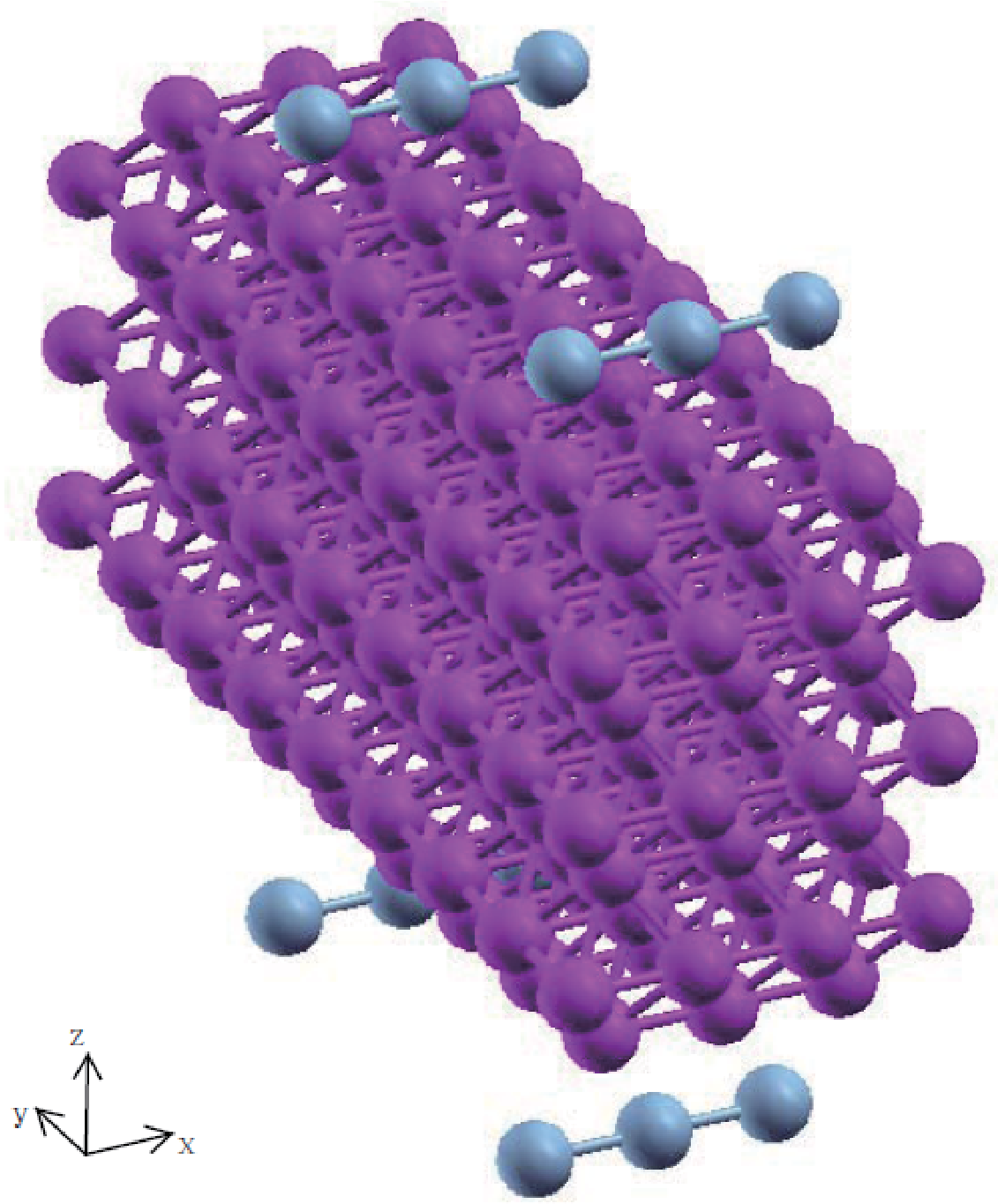}
\caption{(color online) The slab supercell for modeling Fe, Co and Ni nanowires on the hollow sites of Cu (001) surface.}
\label{Cstru}
\end{figure}

In the present calculations, we use the accurate frozen-core full-potential projector augmented-wave (PAW)
method,~\cite{blo94} as implemented in the Vienna {\it ab initio} simulation package (VASP) \cite{vasp1,vasp2}. The calculations
are based on density functional theory with the generalized gradient
approximation (GGA)\cite{Perdew91}. A large plane-wave cutoff energy of 350 eV is used for all the systems considered.
The Fe, Co, and Ni nanowires along the {\it x} direction on the
Cu (001) surface are modeled by a nanowire attached to both sides of a seven-layer-thick Cu (001) slab
as plotted in Fig. \ref{Cstru}. The transition metal (TM) atoms on the nanowires are placed either on the top of surface Cu atoms
[denoted here as the atop (A) site] or at the hollow position on the Cu surface [called here as hollow (H) site].
The two-dimensional unit cell is chosen to be of
$p$(4$\times$1) structure. The nearest in-plane (out of plane) wire-wire distance is larger than
10\AA$ $ (11\AA) which is wide enough to decouple the neighboring wires.
The theoretical lattice constant (3.60~\AA) of bulk copper,
which is in good agreement with experimental Cu lattice constant of 3.61~\AA, is used
as the fixed in-plane lattice constant of the Cu slab. However, the atoms are allowed to move in the out of plane direction,
and the structural relaxations are performed using the conjugate gradient method. The equilibrium structure is obtained when
all the forces acting on the atoms and the stress are less than
0.02 eV/\AA$ $ and 2.0 kBar, respectively. The $\Gamma$-centered Monkhorst-Pack
scheme with a $k$-mesh of $20\times5\times 1$ in the full Brillouin zone (BZ), in conjunction with the
Fermi-Dirac-smearing method with $\sigma = 0.2$ eV, is used to generate $k$-points for the BZ integration.

Because of its smallness, {\it ab initio} calculation of the MAE is computationally very demanding
and needs to be carefully carried out.\cite{guo91} 
Here we use the total energy difference
approach rather than the widely used force theorem to determine the MAE, i.e., the MAE is calculated as the difference in the
full self-consistent total energies for the two different magnetization directions 
concerned. The total energy convergence criterion is 10$^{-8}$ eV/atom. The same $k$-point mesh is used for the
density of states calculations. The MAEs calculated with a denser 32$\times$6$\times$1 $k$-point
mesh hardly differ from that obtained with the 20$\times$5$\times$1 $k$-point mesh
(within 0.02 meV).

\section {Results and Discussion}


The chain formation energy $E^f$ is defined as $E^f =\frac{1}{2}({E_t-nE^{Bulk}_{Cu}-mE^{chain}_{TM}})$
where $E_t$ is the total energy of the system, $E^{Bulk}_{Cu}$ is the total energy
of the bulk Cu, and $E^{chain}_{TM}$ is the total energy of the freestanding transition metal nanowires.
And $n$ and $m$ are the numbers of the Cu and TM atoms in the system, respectively. The calculated
formation energy is listed in Table \ref{table1}. Clearly, the formation energy is lower
when the TM atoms are placed on the hollow sites, as might be expected.
Interestingly, among the Fe, Co, and Ni nanowires, Ni nanowire has the lowest formation energy on
both hollow and atop sites. The ideal interlayer distance for Cu substrate is 1.8~\AA, and from
Table \ref{table1} the interlayer distance between the Fe, Co, and Ni nanowires
and the Cu substrate for hollow (atop) site is 1.64 (2.31), 1.57 (2.27), and 1.55 (2.26)~\AA,
respectively. At hollow (atop) site, the cooper substrate
seems to push (pull) the TM nanowires and the interlayer distance of Ni is smallest (see Table \ref{table1}).
Table \ref{table1} also shows that when the interlayer distance changes, the spin magnetic moment changes
as well. In general, for all considered Fe, Co and Ni nanowires, an increase in interlayer distance
will result in an increase of spin
moment for both sites. The spin moment is larger in the atop site due to the smaller overlap between the
TM nanowires and the Cu substrate.
At equilibrium, the spin moments of Fe, Co and Ni on hollow (atop) sites are 3.07 (3.29), 1.79 (1.99)
and 0.0 (0.65) $\mu_B$/atom.
In our previous study\cite{Tung} of freestanding 3$d$ TM nanowires, the spin moments at the same bond length
(2.55 \AA) of Fe, Co and Ni are 3.30, 2.30 and 1.14 $\mu_B$/atom, respectively.
Clearly, placing the 3$d$ TM nanowires on the hollow sites significantly reduces or even quenches the spin moments
on the nanowires, whilst the spin moments are much less affected when the nanowires are deposited on
the atop sites.

\begin{table}
\caption{Calculated formation energy $E^f$, equilibrium interlayer distance $d_{eq}$, spin magnetic moment per magnetic atom,
$m_s$, and magnetization energy per magnetic atom $E^{mag}$=$E^{FM}$-$E^{NM}$ of the 3$d$ TM nanowires on both the hollow (H)
and atop (A) sites of the Cu (001) surface. Here superscripts $FM$ and $NM$ denote the ferromagnetic and nonmagnetic
states, respectively.}
\begin{tabular}{cccccc}
\\ \hline
      & site & $E^f$& $d_{eq}$ & $m_s$ &$E^{mag}$\\
      &      &  (eV/u.c.) &    (\AA)  & ($\mu$$_B$)& (eV)   \\\hline
   Fe & H &  -2.14& 1.64& 3.07& -0.337\\
      & A &  -0.31& 2.31& 3.29& -0.444\\
   Co & H &  -2.15& 1.57& 1.79& -0.130\\
      & A &  -1.19& 2.27& 1.99& -0.205\\
   Ni & H &  -2.26& 1.55& 0.00&  0.000\\
      & A &  -1.32& 2.26& 0.65& -0.032\\
  \hline\hline
\end{tabular}
\label{table1}
\end{table}

The relativistic SOC is essential for the orbital magnetization and magnetocrystalline anisotropy in solid,
though it may be weak in the 3$d$ TM systems. Therefore,
we include the SOC in our further self-consistent calculations in order to study the magnetic anisotropy and also orbital
magnetization of the Fe, Co and Ni nanowires on Cu (001), and the results are summarized in Table \ref{table2}.
First, when the SOC is taken into account, the spin magnetic moments for the TM nanowires at hollow (atop)
site are 3.07 (3.28), 1.78 (1.99), and 0.02 (0.64) $\mu_B$/mag. atom which are almost identical
to the corresponding one obtained without SOC. This is due to the weakness of SOC in the 3$d$ transition metals.
However, including the SOC gives rise to a significant orbital magnetic
moment in TM nanowires and, importantly, allow us to determine the easy magnetization axis in 3$d$ nanowires.
For the magnetization lies
along the chain direction, the calculated orbital moments for Fe, Co and Ni are 0.10, 0.27 and 0.02 $\mu_B$/mag. atom (see
Table \ref{table2}), respectively. In our previous study of 3$d$ TM nanowires, we found that
the magnetization has a strong directional dependence \cite{Tung} and the orbital moment is larger
when the magnetization lies along the chain direction. In this study, the orbital
magnetic moments for Co and Ni still have a directional dependence but become smaller compared
with freestanding nanowires, as expected, and the orbital magnetization is also larger when the magnetization
is along the chain direction, except Co and Ni on the atop site.
At the atop site, the orbital moment for Ni and Co is larger when the magnetization is in-plane
but perpendicular to the chain direction (i.e., along the $y$-axis).



\begin{figure}
\includegraphics[width=8cm]{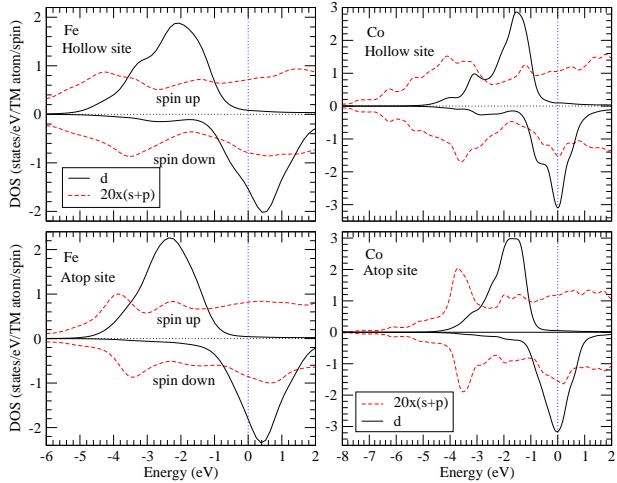}
\caption{(Color online) Spin-polarized density of states of the Fe and Co nanowires on Cu (001).}
\label{Cdos}
\end{figure}

The magnetic moments of the Fe, Co and Ni nanowires can be understood by the calculated spin-polarized density of states (DOS)
as displayed in Fig. \ref{Cdos}. In Fig. \ref{Cdos}, the Fermi levels are set to be zero, the DOS for the
minority spin are multiplied by $-1$, and the $sp$-orbital decomposed DOS are scaled by a factor of 20 for an easy comparison.
Clearly, the $d$-orbitals of the Fe and Co nanowires on both sites are significantly localized due to the reduction
of the coordination number whilst the $sp$-orbitals are more dispersive. The reduction in coordination number also induces
the great enhancement in the spin splitting of Fe and Co 3$d$-bands. The spin-splitting of Fe and Co 3$d$-bands for hollow (atop)
site are 2.58 (2.82), and 1.51 (1.75) eV, respectively. The interlayer distance between the TM nanowires and Cu substrate
are larger on the atop site which means that the overlaps between the TM and substrates are smaller, and the spin magnetic moments
are larger. The splitting of the 3$d$-band is approximately proportional to the spin moment.

The calculated MAEs of Fe, Co and Ni nanowires are listed in Table \ref{table2}.
We define the energy differences $E^1$=$E^{100}$-$E^{001}$ and $E^2$=$E^{100}$-$E^{010}$.
If both $E^1$ and $E^2$ are negative, 
the magnetization prefers to lie along the chain direction. 
First, Co nanowires have the largest MAE on both hollow and atop sites as well as the largest orbital magnetic moment.
Interestingly, both $E^1$ and $E^2$ are positive in the Fe nanowire, indicating that it has an out-of-plane
anisotropy at both sites, whilst the Co and Ni nanowire show an in-plane anisotropy. Recent {\it ab initio} 
studies of Fe double chains on Ir (001) surface\cite{Riccardo09, Mokrousov09} show that the anisotropy 
could be either in-plane or out of plane, depending on the inter-chain distance and the magnetic 
configuration of the chain. 
The MAEs of the freestanding 3$d$ TM atomic chains\cite{Tung} were already obtained by the force theorem method
with a fine 1$\times$1$\times$$n$ ($n$ = 200) $k$-point mesh,  and the calculated  MAE $E^1$ 
was -2.25, 0.68 and -8.13 (meV/mag. atom) for the Fe, Co and Ni chain, respectively. 
For a more precise comparison, 
here we recalculate the MAE of the freestanding Fe, Co and Ni atomic chains by the total energy 
difference approach with $n$ = 200.
The resultant MAE for the Fe, Co and Ni chains is -0.6, 0.84, -8.91 (meV/mag. atom), respectively.
Clearly, when the 3$d$ TM chains were deposited on Cu (001), the MAE generally becomes smaller due
to the overlap between the TM nanowires and substrate (Table \ref{table2}). However, the MAE for Co is enhanced,
though the predicted magnetic anisotropy is in-plane. 
Furthermore, the MAE for the 3$d$ TM nanowires deposited on Cu (001) is still very large when 
compared with that of bulk Fe and Ni.

\begin{table}
\caption{Calculated spin magnetic moment per atom, $m_s$, orbital magnetic moment per magnetic atom, $m_o$, along three different
directions and magnetic anisotropy energy per magnetic atom (MAE) of the Fe, Co and Ni chains on Cu (001). The chain direction
is along (100), and the (010) [(001)] direction is in-plane [out of plane] but perpendicular to the chain direction.
The MAE $E^1$ is defined as $E^{100}$-$E^{001}$ and $E^2$ is $E^{100}$-$E^{010}$.}
\begin{tabular}{ccccccc}
\\ \hline
        & $m_s$               &       &  $m_o$        &      &      MAE  &      \\
        & ($\mu_B$)           &       &  ($\mu_B$)    &      &  (meV)    &      \\
        &                     & 100   &    010        & 001  &   $E^1$   & $E^2$\\ \hline
        &                     &       &    H-site     &      &           &      \\
Fe      &           3.07      & 0.10  &    0.09       & 0.10 &   0.32    &  0.25\\
Co      &           1.78      & 0.27  &    0.18       & 0.17 &  -1.17    & -1.16\\
Ni      &           0.02      & 0.02  &    0.00       & 0.00 &  -0.53    &  0.01\\
        &                     &       &    A-site     &      &           &      \\
Fe      &           3.28      & 0.12  &    0.11       & 0.13 &    0.38   &  0.29\\
Co      &           1.99      & 0.19  &    0.25       & 0.12 &   -0.40   & -1.51\\
Ni      &           0.64      & 0.13  &    0.27       & 0.11 &   -0.05   & -0.32\\
     \hline\hline
\end{tabular}
\label{table2}
\end{table}

\section {Conclusions}
We have performed {\it ab initio} GGA calculations for the Fe, Co and Ni linear atomic chains on Cu (001)
surface in order to examine how the substrates would affect the magnetic
properties of the nanowires. We found that Fe, Co and Ni linear
chains on Cu (001) surface still have a stable or metastable FM state. When the SOC is
included, the spin magnetic moments remain almost unchanged, due to the weakness of SOC in 3$d$ TM chains.
However, the significant
orbital magnetic moments appear and are also direction-dependent, except Fe.
We also found that the Fe system has an out-of-plane magnetic
anisotropy whilst the Co and Ni systems have an in-plane anisotropy.

\section*{Acknowledgments}
The authors acknowledge support from the National Science Council and the NCTS of Taiwan.









\end{document}